\documentclass[sigconf]{acmart}

\AtBeginDocument{%
	\providecommand\BibTeX{{%
			\normalfont B\kern-0.5em{\scshape i\kern-0.25em b}\kern-0.8em\TeX}}}

\usepackage{graphicx}
\usepackage{textcomp}
\usepackage{xcolor}

\usepackage{algorithm}
\usepackage{setspace}
\usepackage{algpseudocode}

\usepackage{multirow}

\usepackage{hyperref}
\hypersetup{colorlinks=true, citecolor=blue}

\usepackage{blindtext}

\begin{document}
	
	
	\title{Solving DWF Dirac Equation Using Multi-splitting Preconditioned Conjugate Gradient with Tensor Cores on NVIDIA GPUs
	}

	\author{Jiqun Tu}
	\affiliation{%
		\institution{NVIDIA Corporation}
		\city{Santa Clara, CA 95051}
		\country{USA}}
	\email{jtu@nvidia.com}

	\author{M. A. Clark}
	\affiliation{%
		\institution{NVIDIA Corporation}
		\city{Santa Clara, CA 95051}
		\country{USA}}
	\email{mclark@nvidia.com}

	\author{Chulwoo Jung}
	\affiliation{%
		\institution{Brookhaven National Laboratory}
		\city{Upton, NY 11973}
		\country{USA}}
	\email{chulwoo@bnl.gov}

	\author{Robert D. Mawhinney}
	\affiliation{%
		\institution{Columbia University}
		\city{New York, NY 10027}
		\country{USA}}
	\email{rdm10@columbia.edu}

	\begin{abstract}
                We show that using the multi-splitting algorithm as a preconditioner for the domain wall Dirac linear operator, arising in lattice QCD, effectively reduces the inter-node communication cost, at the expense of performing more on-node floating point and memory operations.
		Correctly including the boundary \textit{snake} terms, the preconditioner is implemented in the QUDA
                framework, where it is found that utilizing kernel fusion and the tensor cores on NVIDIA GPUs is necessary to achieve a sufficiently performant preconditioner. A reduced-dimension (reduced-$L_s$) strategy is also proposed and tested for the preconditioner.
		We find the method achieves lower time to solution
                than regular CG at high node count despite the
                additional local computational requirements from the preconditioner.
		This method could be useful for supercomputers with more on-node flops and memory bandwidth than inter-node communication bandwidth.
	\end{abstract}

\ccsdesc[500]{Applied computing~Physics}

	\keywords{lattice QCD, domain wall fermion, GPU, tensor core, preconditioned conjugate gradient}
	
	\maketitle
	
	\section{Introduction}
	\subsection{Situation and Motivation}
	
	Quantum Chromodynamics (QCD) is the theory that describes the
        interaction between quarks and gluons, and accounts for most
        of the physical matter we encounter every day.
	Lattice QCD remains as the only viable first-principles method to calculate physical predictions from the non-linear theory.
	However, such calculations are extremely computationally
        demanding: consuming in excess of $10\%$ of total
        cycles at large public supercomputer sites, such as
        NERSC~\cite{deslippe2019}.

	The most computationally expensive part of full lattice
        QCD simulations is the solution of the
        Dirac equation.  On a 4D space-time grid, the discretized
	Dirac equation is a large sparse linear system and solving
        this system is required for both the Markov chain ensemble
        {\it generation} phase, where snapshots of the QCD vacuum are
        produced with the correct Boltzmann weight, and the {\it measurement}
        phase, where the expectation values (or ensemble averages) of
        desired QCD observables are calculated from these
        snapshots. Each snapshot, known as a {\it gauge field}
        configuration gives rise to a unique linear
        system with the Dirac equation.

	Traditionally, Krylov sub-space solvers have been used to solve the Dirac equation and, as detailed below, the conjugate gradient (CG) algorithm~\cite{Hestenes&Stiefel:1952} remains an optimal choice for some stencil discretizations of the Dirac operator.
	The convergence rate is limited by the condition number of the Dirac operator, which has become as large as $10^8$ in current
        simulations due to the small input quark masses required to
        reproduce physical phenomena, e.g., the correct physical pion mass\footnote{See table \ref{tab:info} for condition numbers of the operators used in this work.}.
	Many approaches have been developed to accelerate the
        convergence rate and in this work we describe a method that is
        applicable for solving the Dirac equation for the domain wall
        fermions (DWF) discretization in the ensemble generation phase.
	While this discretization is very appealing from a physics
        point of view, due to its preservation of all of the continuum symmetries, it is this very property that makes the discretization more challenging to solve numerically.
	
	An early step away from a traditional Krylov solver was a
        domain decomposition algorithm proposed for the so-called
        Wilson fermion discretization in~\cite{Luscher2004}.
	This method uses a Schwarz Alternating Procedure (SAP) as a preconditioner for a generalized conjugate residual (GCR) solver.
	This was shown to be faster than the popular BiCGStab algorithm~\cite{10.1137/0913035, Frommer:1994vn}, but a useful implementation
        of this method has not been found for DWF.  The additive Schwarz algorithm has been used for the Dirac equation inversion for various fermions~\cite{Osaki2010, Babich2011},
        but again not for DWF.
        Subsequently, marked improvements in the time to solution have
        been achieved via adaptive multigrid methods~\cite{Brannick:2007ue, Babich:2010qb, Yamaguchi:2016kop}, and deflation/eigenspace methods.
	Deflation/eigenspace methods generally calculate the eigenvectors for low-modes (those with small eigenvalues), and use these to reduce the condition number of subsequent solves.
	For DWF the EigCG algorithm~\cite{Stathopoulos2010} and the
        implicitly restarted Lanczos algorithm with Chebyshev
        polynomial accelerations method~\cite{YSaad1980, Shintani2015}
        have been successfully deployed in current calculations.
        The significant setup costs with both adaptive multigrid and
        eigensolver methods render them more suitable for the
        measurement phase since this requires solving the
        same Dirac equation for many different right-hand sides
        (sources), allowing for amortization of these overheads.  However, this is
        much more challenging for the generation phase where, at most, a half
        dozen linear solves are required per linear system.

	For the ensemble generation phase, the Markov chain Monte
        Carlo method of choice is the Hybrid Monte Carlo (HMC) algorithm~\cite{Duane:1987de}.
	In this algorithm the Dirac equation is solved for each small
        change in the gauge field, and, because the gauge fields are
        changing, each solution is based on a different linear system.
	For Wilson fermions, combining the Schwarz preconditioner with HMC provided an efficient simulation algorithm~\cite{Luscher:2005rx}.
	While there have been various endeavors in applying multigrid
        methods for ensemble generation, for DWF these methods have
        come up short~\cite{McGlynn:2016ahb}, owing to the difficulty
        in breaking even with respect to the setup costs.  The fundamental issue is
        that since the gauge fields are evolving, the setup required for multigrid for one gauge field snapshot becomes less optimal for the next gauge field.
        Additional inherent difficulties arise with DWF, owing to the complex
        indefinite nature of the eigenspectrum of the linear
        operator~\cite{Brower:2020xmc}.\footnote{This contrasts with
          the Wilson discretization which while non-Hermitian,
          generally has a positive real eigenspectrum, allowing for
          direct (non-normal) methods.}

        For multi-node computers, lattice simulations divide the
        global space-time domain into non-overlapping sub-domains that are stored and computed locally on different nodes.
	While this increases the total compute capability, it requires communication between the processors.
	For CG, communication is required for each iteration of
        the solve: 1.) global collectives arising from vector inner
        products, and 2.) halo communication arising from satisfying
        data dependences in the stencil application.  For Lattice QCD
        the latter is the rate limiter, and with which we concern
        ourselves with.  The DWF operator when
        deployed on multi-node computers requires roughly 1 byte of
        local inter-node communication bandwidth for each flop
        (floating point operation) of on-node computing.\footnote{The
          precise requirement varies with the size of the problem on
          each local node.}

        While the architectures of recent
        generations of supercomputers have greatly increased the
        floating point operations per second (flops) on each node, as
        well as the local memory bandwidth, there has not been a corresponding increase in inter-node communication bandwidth.
	On some of the newest machines, for example the SUMMIT machine at ORNL, the communication has become
	the bottleneck: when strong scaling the local computation time is between
        $10$-$20\times$  less than the communication time and the large
        local computational resources are not utilized\footnote{Aggregate single precision flops and total memory bandwidth of the 6 NVIDIA V100 GPUs on a SUMMIT node are 63 TFLOPS and 5.4 TB/s, respectively, while the bi-directional inter-node communication bandwidth is only 50 GB/s.}.
	
	In this work, we report on our investigation into an efficient
        solver for generating ensembles with DWF which consumes {\it more local computation but less communication} than standard CG.
	The cost of the local computation is reduced by an efficient implementation of the solver on NVIDIA GPUs utilizing the tensor cores, and an overall speedup in terms of time to solution is achieved.
	
	\section{Method}
	
	\subsection{Multi-splitting Algorithm}
	
	In~\cite{OLeary1985} the \textit{multi-splitting} algorithm is proposed for solving generic linear systems.
	Compared to SAP, it does not require checker-boarding, using the \textit{current} solution outside of each node as the Dirichlet boundary to perform local solutions.
	This acts as one iteration, and the solution is updated after each iteration, and the new boundary is communicated between nodes to ready the next iteration.
	
	\begin{figure}[]
		\centering
		\includegraphics[width=\linewidth]{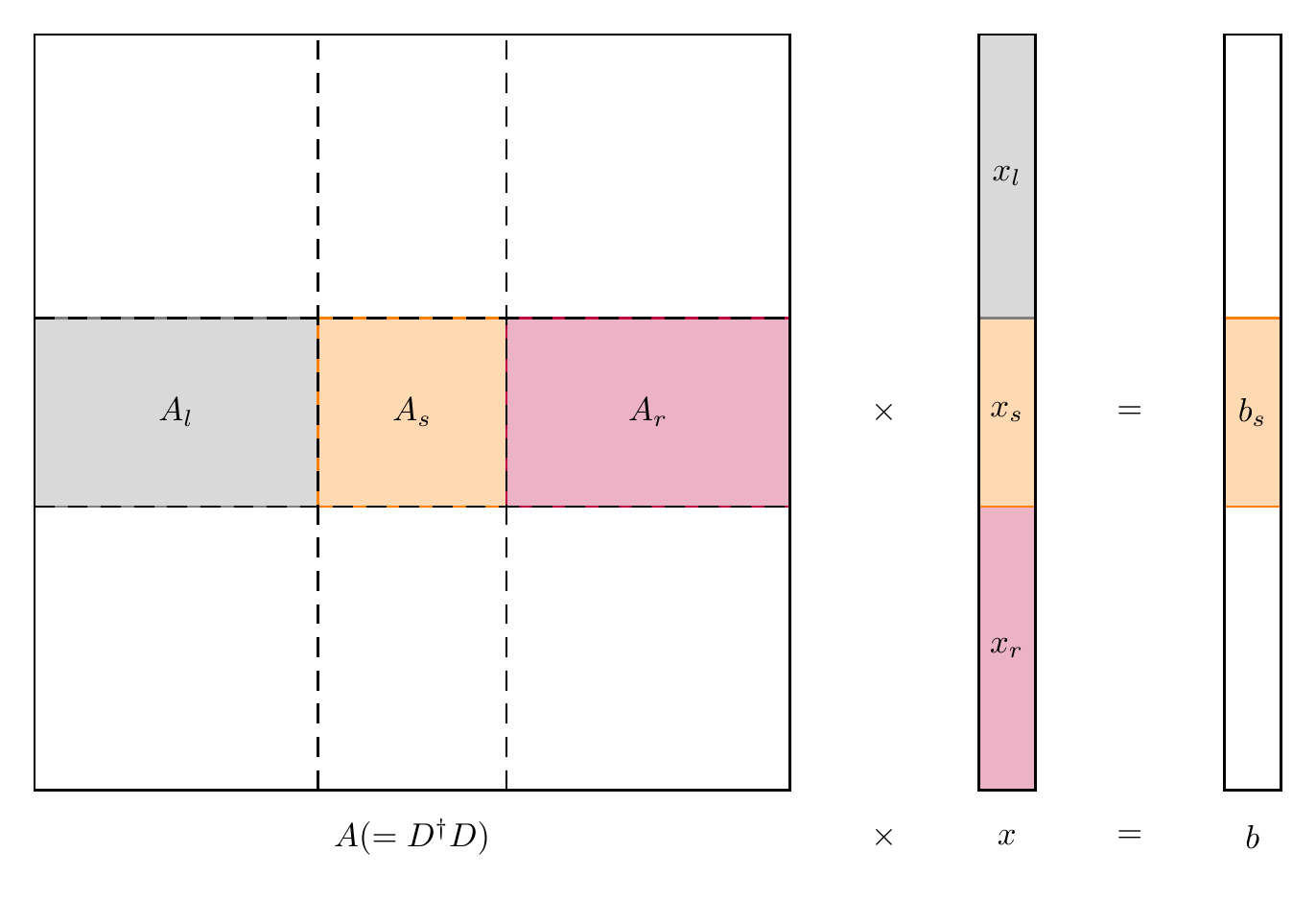}
		\caption{Decomposition of the matrix $A$, the solution vector $x$ and the right-hand-side vector $b$ into local parts on a node.}\label{fig:ms_dec}
	\end{figure}
	Following~\cite{Jezequel2012}, assuming the equation to be
        solved is $Ax=b$, at node $s$ the matrix $A$ and vectors $x$ and $b$ are decomposed according to figure \ref{fig:ms_dec}, where $x_s$ and $b_s$ are the parts that are locally stored.
	The original equation then turns into
	\begin{equation}\label{eq:local}
		A_sx_s+A_lx_l+A_rx_r=b_s.
	\end{equation}
	The $A_lx_l+A_rx_r$ involves off-node vector parts and is calculated before each iteration with inter-node communication.
	Then for each iteration on each node the algorithm solves the equation
	\begin{equation}\label{eq:ms}
		A_sx_s=b_s-A_lx_l-A_rx_r
	\end{equation}
	locally for $x_s$.
	The updated solution $x_s$ will then be communicated to other nodes that need it.
	The local solves could be done concurrently on all nodes once the communication work to calculate $A_lx_l+A_rx_r$ is done.
	
	The domain decomposition SAP algorithm and this algorithm treat the Dirichlet boundary differently and therefore the orders of local inversions are different: the former is multiplicative and requires checker-boarding, while the later is additive and does not require checker-boarding.
	
	\subsection{Domain Wall Fermions}
        \label{sec:dwf}
	The domain wall fermion (DWF)~\cite{Jansen1996} formulation,
        or its variant M\"obius DWF (MDWF)~\cite{Brower2004} which is
        used in this work, is one of several widely used
        discretizations of the continuum 4D Dirac operator. It
        suppresses the breaking of an important physical symmetry of QCD ({\it chiral} symmetry)
        at the cost of adding a fictitious fifth dimension of size
        $L_s$. The MDWF Dirac operator acts on pseudo-fermion vectors,
        which, on each 5D site, contain four \textbf{spin} and three
        \textbf{color} components of complex numbers to make a total
        of 24 real degrees of freedom per 5D site. The spin components
        represent quarks anti-commuting nature as fermions. The
        gauge fields are composed of 4D fields with unique 3-by-3 SU(3)
        matrices between each 4D lattice site.  These matrices act in
        only in color space, i.e., they have no spin or fifth-dimension
        dependence.

	Modern numerical implementations of MDWF utilize the fact that
        only the matrix elements that connect the \textit{even} sites
        to \textit{odd} sites and those connecting \textit{odd} sites
        to \textit{even} sites depend on the gauge field while the matrix elements that connect \textit{even} sites to \textit{even} sites and those connect \textit{odd} sites to \textit{odd} sites do not and are constant.
	Here the even-odd parity is defined by the 4D components of a site:
	\begin{equation}
		\mathrm{parity}\equiv (x+y+z+t)\mod 2.
	\end{equation}
	In this even-odd order form the MDWF Dirac equation can be written as,
	\begin{equation}
		\begin{pmatrix}
			M_5 & M^4_{eo} \\
			M^4_{oe} & M_5 \\
		\end{pmatrix}
		\begin{pmatrix}
			\psi_e \\
			\psi_o
		\end{pmatrix}
		=
		\begin{pmatrix}
			\phi_e \\
			\phi_o
		\end{pmatrix},
	\end{equation}
	where $\psi$ is the desired solution vector, $\phi$ is the
        source vector, the subscript $e/o$ refer to even and odd sites
        and we have suppressed tensor indices for brevity. This is
        equivalent to solving the following even-odd preconditioned
        system (known as red-black Schur complement preconditioning
        in other parlance)
	\begin{align}\label{dirac_equation}
		D\psi_e=\hat\phi_e,\ D&\equiv \mathbf{1}-M_5^{-1}M^4_{eo}M_5^{-1}M^4_{oe}, \\ \hat\phi_e&\equiv M_5^{-1}\phi_e-M_5^{-1}M^4_{eo}M_5^{-1}\phi_o,
	\end{align}
        Here $\mathbf{1}$ is the identity matrix, and $M^4_{eo/oe}$ includes the \textit{Wilson
         -hopping} (4D stencil) term $D^w_{x,y}$ that connects 4D
        space-time sites to their nearest neighbors,
	\begin{align}
		M^4_{oe/eo}&=D^w_{x,y}M_\phi,
	\end{align}
        and $M_5$ and $M_\phi$ are constant matrices that are diagonal
        in the 4D Euclidean space-time dimensions.  The Wilson-hopping
        term is given by
	\begin{align}
		D^w_{x,y}&\equiv\sum_{\mu=1}^4\left[(1+\gamma_\mu)U^\dagger_{x-\hat{\mu},\mu}\delta_{x-\hat\mu,y}+(1-\gamma_\mu)U_{x,\mu}\delta_{x+\hat{\mu},y}\right],
	\end{align}
        and it is here that the underlying gauge-field \(U\) dependence
        comes in: the SU(3) matrices represent the coefficients
        of the nearest-neighbor \(D^w\) stencil.  Since the radius
        \(r=1\) of
        Wilson stencil, each application leads
        to a propagation, or hopping, of a non-zero contribution in each
        dimension and direction of size one.
        Exhaustive details of these
        matrices can be found in~\cite{Brower2004}.   We note that given
        the 4D multi-processor decomposition only the \(D^w\) contribution involves communication between
        processor sub-domains.
	
	The CG algorithm requires the underlying matrix to be
        Hermitian and positive definite, while the matrix $D$ is
        complex indefinite.  Hence Conjugate Gradient Normal Residual (CGNR)
        is often used: both sides of (\ref{dirac_equation}) are left multiplied by the Hermitian conjugate operator $D^\dagger$ and the resulting equation with the normal operator $D^\dagger D$ and the new RHS $D^\dagger\hat\phi_e$ is solved instead,
	\begin{equation}\label{normal_equation}
		D^\dagger D \psi_e=D^\dagger\hat\phi_e.
	\end{equation}

	\subsection{Dirichlet Boundary Condition on the 4-Hop Normal Operator}
	There are four Wilson hopping terms, one in each $M_{eo/oe}^4$, in the normal operator $D^\dagger D$,
	\begin{equation}\label{eq:DdagD}
		D^\dagger D=\big[\mathbf{1}-M_5^{-1}\textcolor{red}{M^4_{eo}}M_5^{-1}\textcolor{red}{M^4_{oe}}\big]^\dagger\big[\mathbf{1}-M_5^{-1}\textcolor{red}{M^4_{eo}}M_5^{-1}\textcolor{red}{M^4_{oe}}\big].
	\end{equation}
	
\begin{figure}[]
		\centering
		\includegraphics[width=\linewidth]{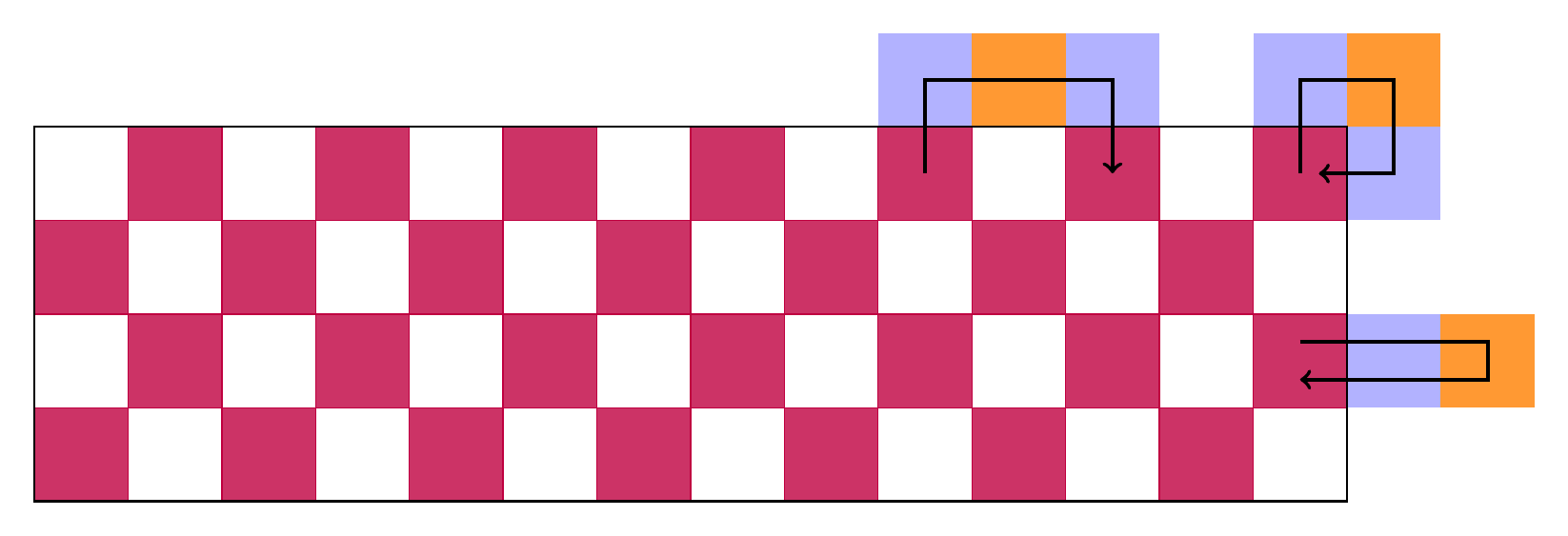}
		\caption{The normal operator $D^\dagger D$ has as many as four Wilson hopping terms (stencils).
			Enforcing Dirichlet boundary condition on it requires the inclusion of the \textit{snake} terms, which are represented, as a 2D illustration, by the black arrows.}\label{fig:snake}
	\end{figure}

	To apply the multi-splitting algorithm to (\ref{normal_equation}) Dirichlet boundary conditions are to be enforced on the normal operator $D^\dagger D$, i.e., the local part (the $A_s$ in (\ref{eq:local})) of this normal operator needs to be constructed.
	As the vector content is distributed across the processors according to its 4D space-time location, this local part for $D^\dagger D$ includes \textit{snake} terms that hop out of the boundary and hop back in as the various components in (\ref{eq:DdagD}) are evaluated.
	Figure \ref{fig:snake} illustrates this and gives some examples of the snake terms.
	These terms are truncated if Dirichlet boundary conditions are enforced on each of the four $M^4_{eo/oe}$ hopping terms sequentially. Using $\lfloor\bullet\rfloor$ to indicate applying Dirichlet boundary conditions,
\begin{equation}\label{eq:diri}
  \lfloor M^4_{eo}\rfloor \lfloor M^4_{oe}\rfloor \lfloor M^4_{eo}\rfloor \lfloor M^4_{oe}\rfloor \neq \lfloor M^4_{eo} M^4_{oe} M^4_{eo} M^4_{oe}\rfloor.
\end{equation}
	Our simulation results show that the correct inclusion of these snake terms is crucial to the convergence.
	
	\subsection{Multi-splitting Algorithm as a Preconditioner of CG}
	In~\cite{Luscher2004} to achieve faster convergence the domain decomposition algorithm is eventually used as a preconditioner of GCR.
	In this work, since we are solving an Hermitian system, the multi-splitting algorithm is used as a preconditioner of CG.
	
	Pseudo-code for a generic preconditioned CG\footnote{For a reference of the preconditioned CG algorithm, see \url{https://en.wikipedia.org/wiki/Conjugate_gradient_method}.} is shown in algorithm \ref{alg:cg}, where we are solving $Ax=b$ and $M$ is the preconditioner.
	The preconditioning step is marked with violet background.
	The overall convergence rate of preconditioned CG is estimated by the condition number of $AM^{-1}$. If the condition number of $AM^{-1}$ is smaller then that of the original matrix $A$, faster convergence rate will likely be achieved.
	\begin{algorithm}
		\setstretch{1.15}
		\caption{Preconditioned Conjugate Gradient $Ax=b$}
        \label{alg:cg}
		\begin{algorithmic}
			\State ${r}_0 = {b} - {A x}_0$
			\State ${z}_0 = {M}^{-1} {r}_0$
			\State ${p}_0 = {z}_0$
			\State $k = 0$
			\While {have not converged}
			\State $\alpha_k = {\langle{r}_k,{z}_k\rangle}/{\langle{p}_k,{A p}_k \rangle}$
			\State ${x}_{k+1} = {x}_k + \alpha _k {p}_k$
			\State ${r}_{k+1} = {r}_k - \alpha _k {A p}_k$
			\State \colorbox{blue!30}{${z}_{k+1} = {M}^{-1} {r}_{k+1}$}
			\State $\beta _k = {\langle {z}_{k+1}, {r}_{k+1}\rangle}/{\langle {z}_k,{r}_k \rangle}$
			\State ${p}_{k+1} = {z}_{k+1} + \beta _k {p}_k$
			\State $k = k + 1$
			\EndWhile
		\end{algorithmic}
	\end{algorithm}
	
	Now for this preconditioning step we use the multi-splitting algorithm to solve for $z_{k+1}$ in
	\begin{equation}
		Az_{k+1}=r_{k+1}.
	\end{equation}
	To avoid inter-node communication, a zero initial guess ($x_l=x_r=0$) is used in (\ref{eq:ms}) and only the first iteration is performed.
	With $r_{k+1}$ as the RHS and $z_{k+1}$ the solution,
	\begin{equation}
		A_s x_s = b_s -A_lx_l - A_r x_r \rightarrow A_s z_{k+1,s} = r_{k+1,s}.
	\end{equation}
	This is equivalent to using the local part of the matrix $A$, $A_s$, on each processor as the preconditioner $M$ in the preconditioned CG,
	\begin{equation}
		M=\bigoplus_s A_s,\ s=\mathrm{node\ index}.
	\end{equation}
	The local nature of $A_s$ makes it possible to perform the preconditioning step concurrently on all nodes without communication.
	We refer to this as multi-splitting preconditioned CG (MSPCG).
	
	We note that while the multi-splitting algorithm can split the general matrix $A$ in a variety of ways, the splitting presented here, used as a preconditioner in CG, makes it equivalent to the additive Schwarz algorithm.
	We use the name MSPCG, as it is through the process of applying the multi-splitting algorithm to the MDWF Dirac equation that we realized the necessity of including the snake terms in the local matrix.
	
	\subsection{The Effect of the Preconditioner on the Iteration Count}
	
	The multi-splitting preconditioned CG is applied to solve Dirac equations on three lattice ensembles generated with M\"obius domain wall fermions, using full lattice QCD with the quark masses set at their physical input values.

        Regular CG is used to perform the linear solve in the preconditioning step.
	Instead of adopting a precision-based stopping condition, a fixed number of CG iterations, which will be referred as \textit{inner iterations}, are performed for these preconditioning solves.
	The iterations performed in the overall preconditioned CG will be referred as \textit{outer iterations}.
	In table \ref{table:result} the number of outer iterations needed for the preconditioned CG to converge are reported on the different lattice ensembles, together with the stopping condition for the outer CG (precision) and the processor grid size used.
	The numbers of iterations to reach the same precision with standard CG are also included for comparison, where the inner iteration number is marked with \textit{plain}.
	
	\begin{table*}[t]
		\renewcommand{\arraystretch}{1.5}
		\centering
		\caption{Number of outer iterations need to converge the multi-splitting preconditioned  CG for the lattice ensembles tested in this work.
			\textit{Inner iterations} refers to the fixed number of CG iterations performed for the preconditioning inversion.
			Rows marked with \textit{plain} indicate the iteration count for the same standard CG to converge.
    d.o.f. refers the number of real degrees of freedom of the linear system.}\label{table:result}	
		\begin{tabular}{ccccccc}
			\hline
			\hline
			lattice 4D volume & $L_s$ & d.o.f. & solver tolerance & processor grid size & inner iterations & outer iterations \\
			\hline
			\hline
			\multirow{4}{*}{$32^3\times 64$} & \multirow{4}{*}{$12$} & \multirow{4}{*}{$6.0\times 10^8$} & \multirow{4}{*}{$10^{-8}$} & \multirow{4}{*}{$2^3\times 4$} & plain & $13594$ \\
			&&& & & $3$ & $9106$ \\
			&&& & & $4$ & $6020$ \\
			&&& & & $6$ & $5126$ \\
			\hline
			\hline
			\multirow{4}{*}{$64^3\times 128$} & \multirow{4}{*}{$10$} & \multirow{4}{*}{$8.0\times 10^9$} & \multirow{4}{*}{$10^{-10}$} & \multirow{4}{*}{$4^3\times 8$} & plain & $18092$ \\
			&&&&  & $6$ & $6008$ \\
			&&&&  & $12$ & $5083$ \\
			&&&&  & $18$ & $4948$ \\
			\hline
			\hline
			\multirow{2}{*}{$80^2\times96\times 192$} & \multirow{2}{*}{$32$} & \multirow{2}{*}{$1.1\times 10^{11}$} & \multirow{2}{*}{$10^{-10}$} & \multirow{2}{*}{$4^2\times 8^2$} & plain & $16783$ \\
			&&&& & $6$ & $5719$ \\
			\hline
			\hline
		\end{tabular}
		
	\end{table*}
	
	We have two observations in terms of the effect of the preconditioner:
	
	\begin{enumerate}
		\item Typically on these ensembles with $6$ inner iterations the preconditioned CG reduces the outer iteration count by a factor of $3$.
		Beyond this, while more inner iterations can reduce the outer iteration
                count, there is only asymptotic improvement. This
                saturation is not surprising since increasing the
                inner iteration count is simply solving the
                preconditioning system with increasing accuracy, beyond which no further numerical benefit can be exploited.
		\item Executing a fixed number of inner CG iterations for the preconditioning inversion, instead of using a precision-based stopping condition, does not jeopardize the convergence of the outer CG despite the fact that CG, in principle, is not a flexible solver.
		This is true even when as few as $3$ inner iterations are performed.
		This has previously been observed in~\cite{Golub1999}.
	\end{enumerate}
	
	Our results show that MSPCG reduces the number of outer iterations needed to solve the MDWF Dirac equation, reducing the inter-processor communication at the expense of performing local inner iterations for each outer iteration.
	The local inner iterations need to be sufficiently cheap compared to the outer matrix multiplication (the application of the matrix $A=D^\dagger D$) and corresponding BLAS (Basic Linear Algebra Subprograms) operations in order to achieve an overall speedup in terms of the time to solution.
	
	As a consequence of this and the two observations, the inner iteration count is a parameter that can be tuned to achieve maximum speedup in the trade-off between inter-processor communication and local computation.
	
	\section{Implementation of the Method on NVIDIA GPUs in QUDA}
	The solver is implemented in QUDA, an open source library for
        performing calculations in lattice QCD on graphics processing
        units (GPUs), leveraging NVIDIA's CUDA platform
       ~\cite{Clark:2009wm}.  With QUDA, we are able to fully utilize
        the global memory bandwidth, the different levels of cache,
        and, unique with this work, utilize the tensor cores
        available on recent NVIDIA GPUs, to achieve speedup.
	
	\subsection{An Efficient Implementation of the Preconditioner in QUDA with Tensor Core}

        \subsubsection{Efficient Stencil Application}
        As with other stencils acting on a regular grid, the most
        efficient implementation is generally obtained using a {\it
         matrix-free} formulation, encoding the action of the stencil
        on a grid, as opposed to an implementation as an explicit
        sparse-matrix-times-vector formulation.  It is this approach
        that QUDA employs for all of the stencils it supports.  With an arithmetic
        intensity of around 1--2 in single precision, it is a straightforward roofline
        analysis to see that an efficient implementation of the MDWF
        stencil will be memory-bandwidth bound on
        a single GPU.

        Within a single node, with multiple GPUs, the bandwidth provided by
        the NVLink interconnect is ample to allow near perfect scaling over a node,
        however, when strong scaling on large-scale supercomputers,
        the performance of the matrix $A=D^\dagger D$ is typically
        limited by network bandwidth.  This leaves little room for
        improvement from a software point of view without first
        innovating with respect to the choice of algorithm.

	\subsubsection{Capturing the Snake Terms}
	
\begin{figure}[]
		\centering
		\caption{A 2D illustration of using padded field to capture the snake terms. The orange and purple squares represent non-zero sites, and the white space represent zero sites. Black arrows represent the stencils relevant to the Dirichlet boundary condition. The stencil operators are not applied to the grey areas as they are either zero (as in 2)) or not needed for the output field at the center (as in 5) and 6)).}
        \label{fig:dw}
		\includegraphics[width=\linewidth]{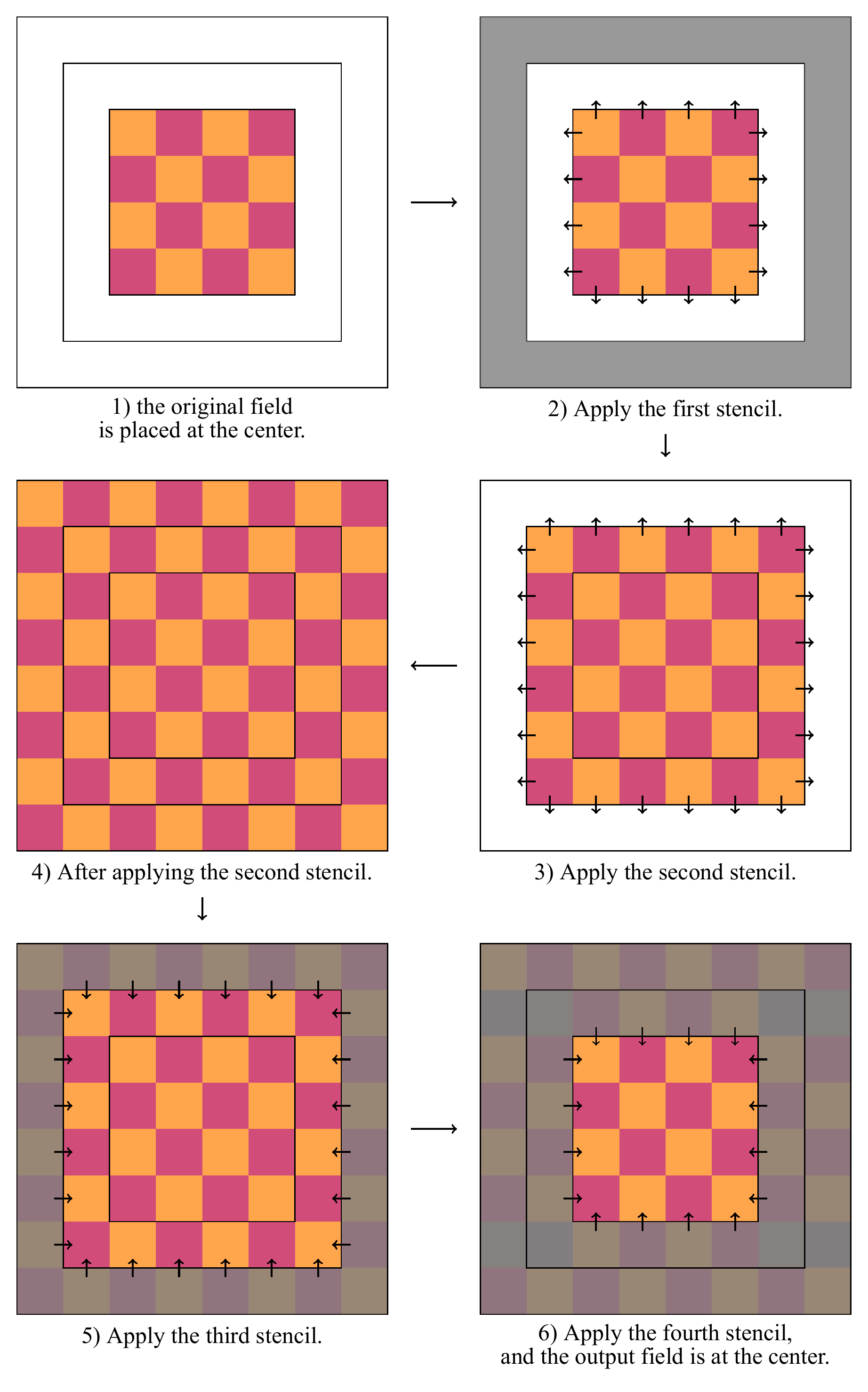}
	\end{figure}

	To apply the preconditioner, the snake terms must be included in order to enforce the correct Dirichlet boundary condition on (\ref{eq:DdagD}).
	In our implementation the vector is padded in both directions
        of each partitioned dimension by size of two and the original field is placed at the center of the padded field while the pad region is initialized to zero.
	The operators in (\ref{eq:DdagD}) of the full normal operator are then applied to the padded vector sequentially with zero Dirichlet boundary condition on the padded boundary.
	The four successive hops are propagated through the padded region.
	The padding size of two is sufficient, since with only four hops, in order to go back to the center region the snake terms are only allowed to hop out of the center region by as far as two steps. A 2D illustration is shown in figure \ref{fig:dw}.

	This approach reuses the existing code for calculating the
        hopping term and the already highly optimized parallelizing
        strategies in QUDA. In order to avoid unnecessary work, we
        only update the minimum number of sites required to ensure
        correct evaluation; this means that a different number of
        sites are updated depending on which stage of the
        computation we are performing.  For example, at the end of the
        computation, where we only desire the final result in the
        center region, there is no point updating the padded region.

	By applying the hops successively the later hops reuse the results of the previous hops, which increases computation reuse compared to applying the individual 4-hop snake terms separately.
	
	\subsubsection{Kernel Fusion}
	
	A closer look at (\ref{eq:DdagD}) gives the following series of operators
	\begin{equation}\label{eq:detail}
		[ \mathbf{1}-M_\phi^\dagger D^{w\dagger} M_5^{-\dagger} M_\phi^\dagger D^{w\dagger} M_5^{-\dagger} ][ \mathbf{1}-M_5^{-1}D^w M_\phi M_5^{-1} D^w M_\phi].
	\end{equation}
	Mathematically the operators are matrices to be left multiplied on the vector sequentially; numerically the operators are functions to be called to apply these multiplications; on the GPUs
	the operators are implemented as kernels to be launched.
	Conceptually each kernel carries out these three steps:
	\begin{itemize}
		\item Load the input vector from GPU memory into the registers;
		\item Perform computation with the data in the registers;
		\item Store the output vector from the registers to GPU memory.	
	\end{itemize}
	
	Skipping the $[\mathbf{1}-\cdots][\mathbf{1}-\cdots]$ part in (\ref{eq:detail}) for simplicity, we have 12 operators to be applied sequentially.
	Input of the kernels is exactly the output of the previous ones.
	Instead of storing the output all the way down to memory and then loading the very same data all the way up to the registers, a better strategy is to keep all the data in cache to avoid the memory traffic and latency from the repeated loading and storing from global memory.
	While this cache reuse would be automatically done by the GPUs
        in the L2 cache if the total memory footprint is less than the
        L2 cache size, the working set size of the problem at hand far
        exceeds the L2 capacity.	In this work, data reuse is
        achieved by utilizing the structure of the matrices and the
        \textit{shared memory}, utilizing explicit kernel fusion to
        increase the arithmetic intensity of the kernels.

	The $D^w$'s in (\ref{eq:detail}) are matrices that hop in 4D
        space-time dimensions and are diagonal in the fifth dimension
        (4D stencil operators); the $M_\phi$'s and $M_5^{-1}$’s are
        matrices that only operate in the fifth dimension and are
        diagonal in the 4D space-time dimensions (5D operators).  As a
        result we can fuse an arbitrary number of 5D
        operators after a given application of \(D^w\), so long as the
        extent of the fifth dimension is contained completely with in
        a single thread block, and hence can be communicated
        through shared memory.  Note
        that fusing multiple $D^w$'s into one kernel is not
        feasible since it requires synchronization between
        thread blocks.
	
	The kernels are deployed as a 2D grid of threads, with the
        {\bf x} dimension corresponding to the 4D space-time points,
        and the {\bf y} dimension corresponding to the 5D index.  For
        each thread block the 4D operator \(D^w\) is first applied and
        the results stored in shared memory; for the subsequent 5D
        operators, all of their input requirements will
        be what is held in the shared memory and so kernel fusion can
        be readily applied.  The size of the thread block
        is chosen such that there is sufficient shared memory to
        hold the intermediate data, with autotuning of the block size
        further deployed to maximize the L2 hit rate for the initial 4D operator application.
	
	Instead of launching 12 kernels for the 12 matrices in (\ref{eq:detail}) only the following 5 fused kernels are launched:
	\begin{equation}
		M_\phi^\dagger D^{w\dagger}, \; M_5^{-\dagger} M_\phi^\dagger D^{w\dagger},\; M_5^{-\dagger} M_5^{-1}D^w,\; M_\phi M_5^{-1} D^w,\; M_\phi.
	\end{equation}

        \subsubsection{Tensor Core}
	
	The $M_5^{-1}$ operator in (\ref{eq:detail}) is a dense matrix
        multiplication in the fifth dimension, and the number of
        floating point operations needed scales as
        $\mathcal{O}(L_s^2)$ for a parallel implementation.\footnote{An $\mathcal{O}(L_s)$ algorithm exists for a
          sequential implementation.} The computation required for
        the $M_5^{-1}$ application could become a bottleneck in the
        fused kernels, which are no longer trivially memory-bandwidth
        bound.  This was indeed what we found in our early
        experiment, where we found
        that the preconditioner was rendered too expensive to improve
        the actual time to solution on the Piz Daint supercomputer
        (featuring the Pascal architecture).

\begin{figure*}[]
		\centering
		\caption{An illustration of how the $M_5^{-1}$, with $L_s=12$, and the number of 4D space-time sites dealt with in the CUDA thread block (``4d'') being 16, is applied using the HMMA instruction. The matrix multiply is of shape ``$48$-by-$48$'' $\times$ ``$48$-by-$96$'' $=$ ``$48$-by-$96$'', and is performed in the sub-block form of ``$16$-by-$16$'' += ``$16$-by-$16$'' $\times$ ``$16$-by-$16$'' with tensor cores as shown in the figure. Note that sizes of the shapes shown here are not to scale.}
        \label{fig:hmma}
		\includegraphics[width=\linewidth]{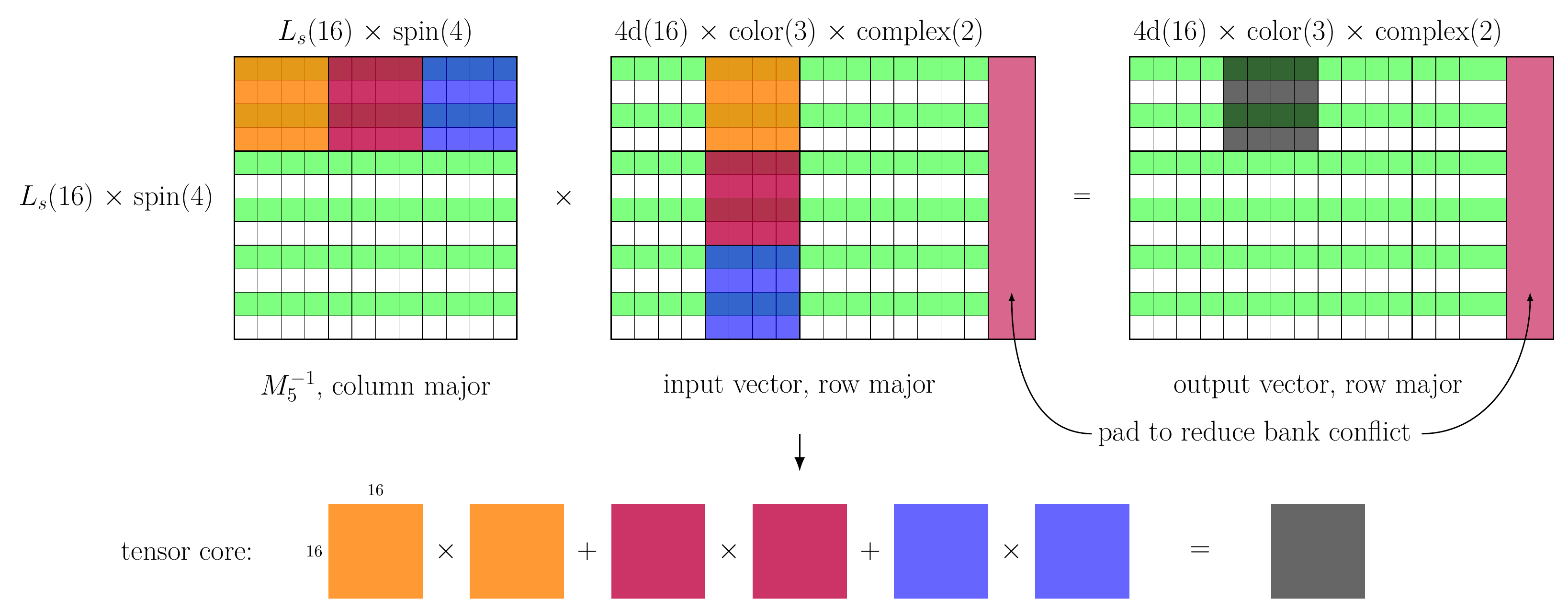}
\end{figure*}

	As a remedy to this problem, the tensor cores and the HMMA
        (Half-precision Matrix Multiply and Accumulate) instructions,
        available on the NVIDIA Volta, Turing and Ampere GPUs, are
        used to speed up the application of this operator.  Following
        the kernel fusion strategy described in the previous part the
        intermediate data is held in shared memory before applying
        $M_5^{-1}$.  Since no global memory loads are required, this
        allows the HMMA instruction to be applied without
        being amortized by any global memory operation latency. As an illustration we show the how the HMMA instruction is applied in the $L_s=16$ case in figure \ref{fig:hmma}.
	
	The half precision (IEEE FP16) has a much narrower representation range ($6.1\times 10^{-5}$ to $65504$) than that of single precision (IEEE FP32, $1.18\times 10^{-38}$ to $3.4\times 10^{+38}$).
	In order to avoid overflow and underflow, the maximum absolute value of the all the numbers in the thread block is found and the numbers are scaled such that they are representable in half precision before the HMMA instructions are applied.
	After the matrix multiplications the numbers are scaled back before being stored to memory as output.
	With a highly parallelized reduction and scaling code this overhead is negligible compared to the overall computation.

	\subsection{Reduced Precision Strategies of Preconditioned CG}
	
	QUDA supports 8-bit and 16-bit fixed-point formats as part of the effort to speed up memory-bandwidth limited kernels.
	For these kernels the floating point numbers are stored as 8-bit and 16-bit signed integers, together with a single-precision floating point number as the scale for every color-spinor ($\textbf{color} (3) \times \textbf{spinor}(4) \times \textbf{complex} (2)=24$ real numbers):
	\begin{equation}
		\textbf{floating point} = \frac{\textbf{integer}}{\textbf{integer-limit}}\times \textbf{scale}
	\end{equation}
	In the kernels these fixed-point numbers are unpacked into single precision floating point numbers. Once they are loaded into registers the computations are done in single precision.
	
	In our simulation most of the outer preconditioned CG iterations are performed with lower precision (16-bit fixed-point format) matrix multiplication and BLAS operations.
	For the solution to achieve higher precision (double precision) a reliable update strategy is deployed: the floating point error accumulated in lower precision during the lower-precision iterations is corrected with higher-precision matrix multiplication and BLAS operations when a certain condition (detailed in~\cite{VanderVorst2000}) is reached.
	
	For the inner preconditioner itself, the matrix multiplication and BLAS operations use the 8-bit and 16-bit fixed-point formats. For the fused tensor-core kernels the numbers are first loaded in fixed-point formats, unpacked to single-precision floating point numbers for applying the stencils, then scaled and converted to half-precision floating point for applying the tensor-core instructions before being packed and stored in fixed-point format. Preconditioning kernels with 8-bit fixed-point format runs faster due to its lower memory traffic but there is a noticeable impact on the quality of the preconditioner, leading to increased outer iterations compared to the 16-bit case. The choice between 8-bit and 16-bit preconditioner depends on their overall effect on time to solution.

	The inexact nature of lower precision also necessitates the use the of the Polak-Ribi\`ere\footnote{See \url{https://en.wikipedia.org/wiki/Conjugate_gradient_method}.} version of the $\beta_k$ update in the preconditioned CG, i.e.,
	\begin{equation}
		\beta _k = {\langle {z}_{k+1}, {r}_{k+1}-r_{k}\rangle}/{\langle {z}_k,{r}_k \rangle}.
	\end{equation}

\subsection{Accelerate the preconditioner with reduced-$L_s$ operators}
In the previous sections we have described our efforts on minimizing the cost of applying the local preconditioner on the technical side by efficiently utilizing the memory and cache bandwidth, as well as the tensor cores. In this section we detail our attempts to minimize the preconditioning cost on the algorithm side.

The kernels launched as part of the preconditioning are
predominantly either global-memory-bandwidth bound, or
L2-cache-bandwidth bound for GPUs with large cache\footnote{For
  example, the Ampere A100 GPU has a large L2 cache of 40 MiB.}. The
time spent on the execution of these kernels is roughly proportional
to the local 5D volume of the underlying vector. Thus, the cost of the
preconditioning solve could be reduced if the fictitious fifth
dimension $L_s$ is shrunk, and if this reduced-$L_s$ operator still
captures the essential spectral qualities of the preconditioner, an overall
speedup could be achieved. Such an approach would be a simplified two-grid algorithm applied in the fifth dimension.

The idea of using reduced-$L_s$ MDWF operators to accelerate
the linear solvers has been explored in the M\"obius accelerated
domain wall fermion (MADWF) algorithm~\cite{Yin:2012rX}, where the
domain-overlap equivalence~\cite{BROWER2006191} has been used to
transform between domain wall systems with different $L_s$. In our
case, however, this approach inflicts too high a numerical cost, since
MADWF requires additional nested linear solves.

Our simplified approach proceeds as follows. Assuming the original Dirac operator with the original $L_s$ is $M$ (and the Dirac operator with reduced-$L_s$, denoted as $\hat{L}_s$, is $\hat{M}$) the original preconditioning step in preconditioned CG is
\begin{equation}\label{.}
  z = M^{-1}r.
\end{equation}
Note that the inversion in $M^{-1}$ is performed with CG with a small fixed number of iterations. It is to be replaced with the reduced-$L_s$ version
\begin{equation}\label{.}
  z = (T^\dagger \hat{M}^{-1} T +\mu \cdot\mathbf{1}) r,
\end{equation}
where the matrix $T$ transforms a vector with $L_s$ into a vector with $\hat{L}_s$, and the corresponding matrix $T^\dagger$ transforms a vector with $\hat{L}_s$ into a vector with $L_s$. This transform matrix $T$ acts on the vector as
\begin{equation}\label{.}
  [Tv]_{s,\mu}=\sum_{t, \nu}[T]_{st,\mu\nu}[v]_{t,\nu},\, t=1,2,\cdots,L_s,\, s=1,2,\cdots,\hat{L}_s,
\end{equation}
where $\mu$,$\nu$ are spin indices. $T$ determines the transformation between different $L_s$'s and a \textit{good} $T$ reduces the cost of applying the preconditioning inversion while still managing to reduce the outer iteration count. The matrix $T^\dagger \hat{M}^{-1} T$ is Hermitian but not necessarily positive definite: zero modes exist since its rank is strictly smaller than the dimension of the vector $r$, if $\hat{L}_s<L_s$. The additional factor $\mu\cdot\mathbf{1}$ is inserted to suppress these zero modes, and its appearance is necessary to make the reduced-$L_s$ preconditioned CG converge.

In our approach $T$ is determined with a method that shares the idea of machine learning: the matrix $T$ is trained such that it minimizes the loss function of
\begin{equation}\label{.}
  \chi^2=\sum_i |M \cdot (T^\dagger \hat{M}^{-1} T+\mu \cdot\mathbf{1})\cdot v_i-v_i|^2,
\end{equation}
where the $v_i$'s are a set of vectors that have significant share of
low-mode components of $M$. In our simulations the $v_i$ are the
residual vectors $r_k$ in a preconditioned inversion with the original
operators with the original $L_s$. In the case where $L_s=\hat{L}_s$,
$M$ would be the same as $\hat{M}$ and the optimal $T$ would be the
identity matrix, and the optimal $\chi^2$ would be zero. With the cost
function $\chi^2$ defined and the its derivatives with respect to $T$
analytically available, $T$ is trained with the line search
method\footnote{For a reference of the line search method, see
  \url{https://en.wikipedia.org/wiki/Line_search}.} to minimize
$\chi^2$. The minimization convergence is accelerated with the momentum method~\cite{pmlr-v28-sutskever13}.

The value of the suppression factor $\mu$ is predefined for the minimization process and its choice is critical to the time-to-solution speed up of the algorithm. $\mu$ is tuned such that the resulting trained $T$ reduces the most outer CG iteration count ($\mu$ being too large makes $T^\dagger \hat{M}^{-1} T+\mu \cdot\mathbf{1}$ close to an identity matrix) while is still able to suppress the zero modes of $T^\dagger \hat{M}^{-1} T$.

The transform matrix $T$ is only non-trivial in the spin components and the fifth dimension coordinates; and is trivial (diagonal) in the color degrees of freedom.  Thus the $T$ trained on a single gauge field configuration can be used to accelerate linear solves performed on other configurations of the same ensemble, which amortizes the cost of training $T$.

	\section{Results}
\begin{figure*}[]
		\centering
        \caption{\(L_2\) residual as a function of (outer) iteration number for MSPCG (with and without reduced-$L_s$ acceleration) used on Dirac equation ($D^\dagger Dx=y$) linear solves on SUMMIT to a tolerance of $10^{-12}$ on the \texttt{RBC96} lattice. $y$ is a Gaussian random source vector.}
        \label{fig:residual-rbc}
		\includegraphics[width=\linewidth]{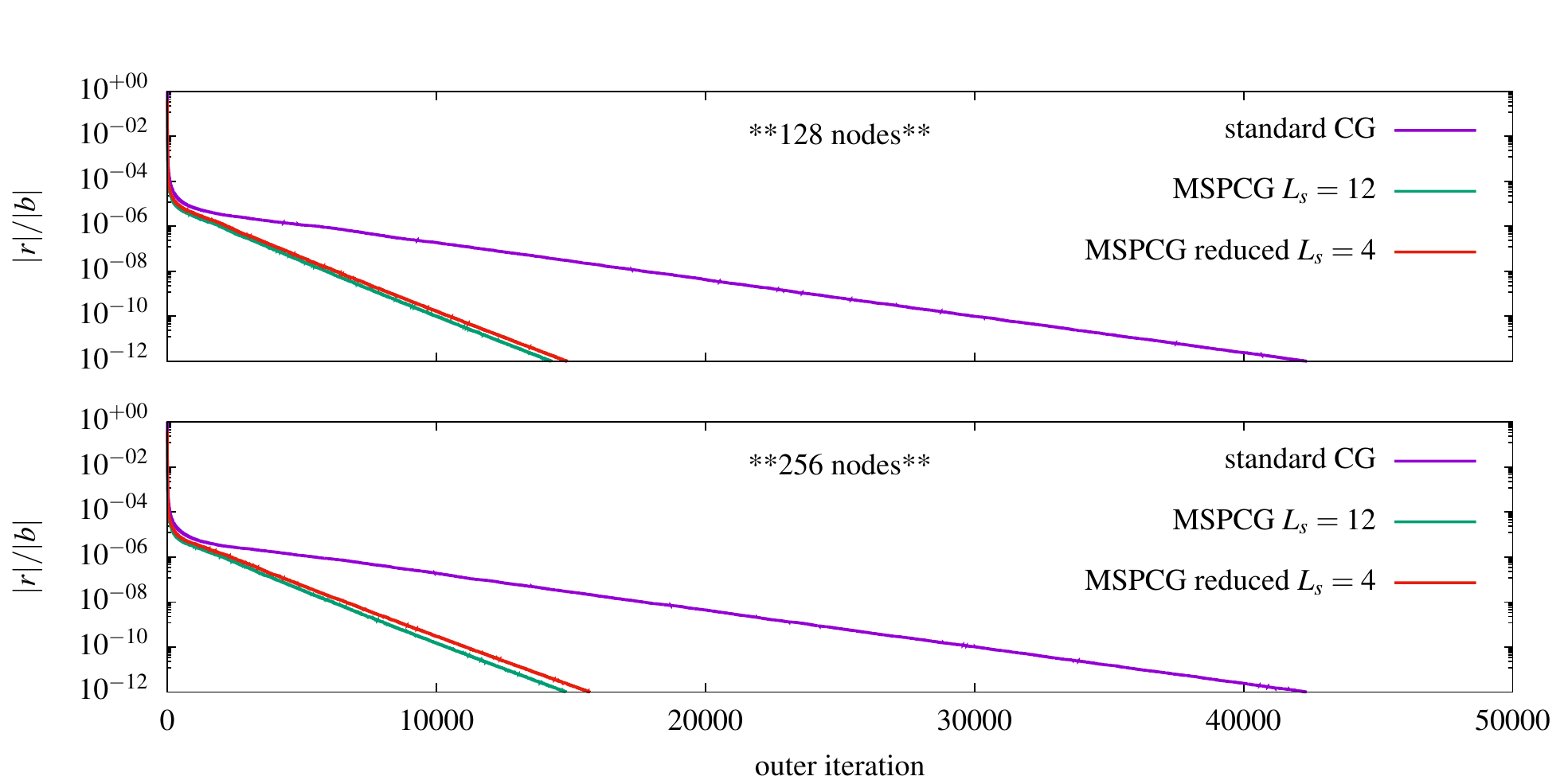}
	\end{figure*}

\begin{figure*}[]
		\centering
        \caption{\(L_2\) residual as a function of (outer) iteration number for MSPCG (with and without reduced-$L_s$ acceleration) used on Dirac equation ($D^\dagger Dx=Dy$) linear solves on SUMMIT to a tolerance of $10^{-10}$ on the \texttt{CAL64} lattice. $y$ is a random source vector.}
        \label{fig:residual-cal}
		\includegraphics[width=\linewidth]{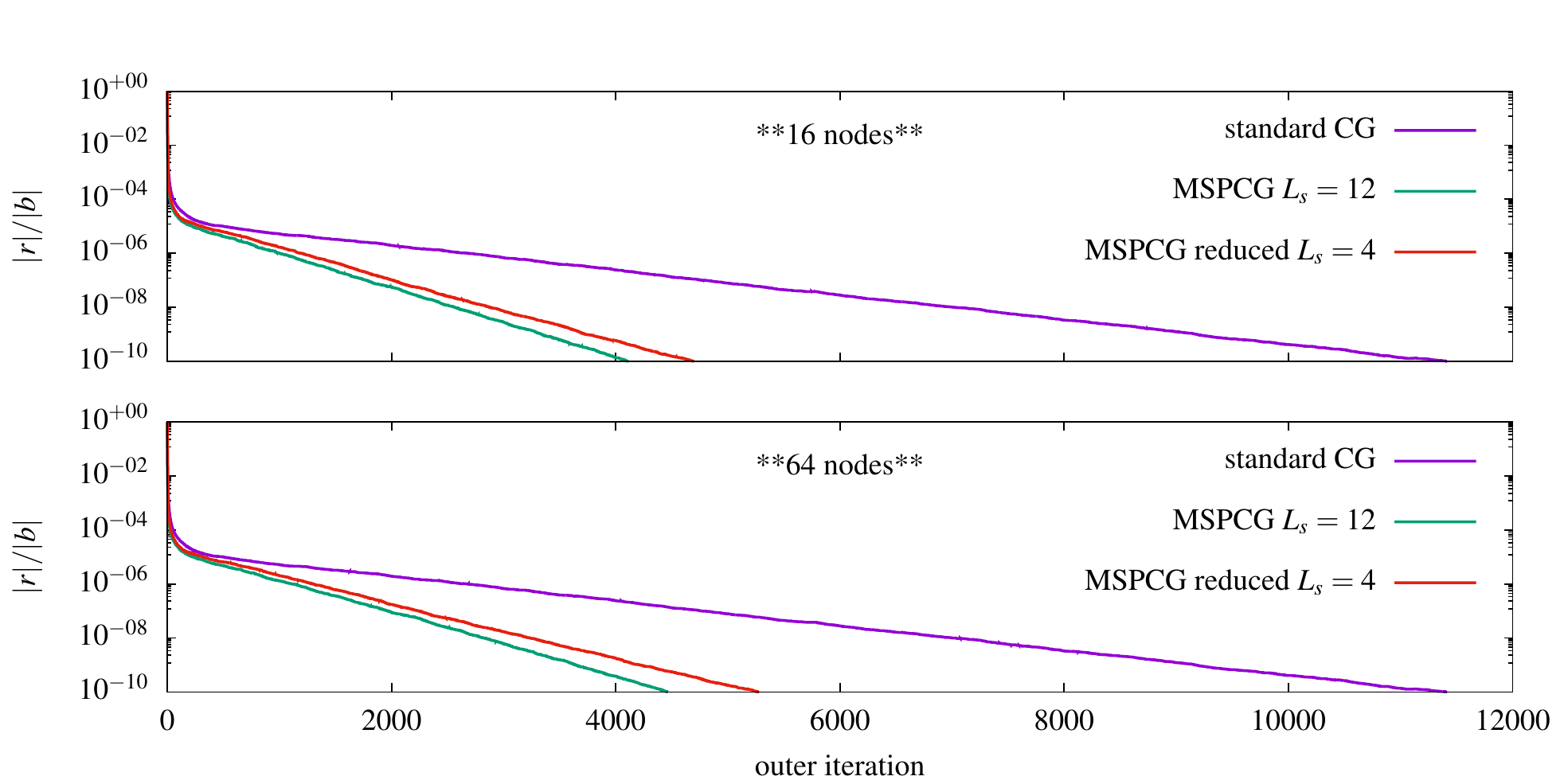}
	\end{figure*}

The multi-splitting preconditioned CG, together with the reduced-$L_s$ (machine learning) technique, is applied to the following two cases, both on SUMMIT:
\begin{itemize}
  \item Solving MDWF Dirac equations for the light quark determinants
    in the evolution phase of the RBC/UKQCD collaborations'
    \texttt{RBC96} lattice. The results are shown in figure
    \ref{fig:residual-rbc} and table \ref{tab:rbc}. On $256$ SUMMIT nodes MSPCG is able to achieve a $51\%$ speedup over standard CG.

  \item Solving MDWF ($L_s=12$) Dirac equations for the light quark correlators in the measurement phase of the CalLat collaborations' \texttt{CAL64} lattice~\cite{Miller:2020xhy,Miller:2020evg}. The results are shown in figure \ref{fig:residual-cal} and table \ref{tab:cal}. On 64 SUMMIT nodes MSPCG is able to achieve a 31\% speedup over standard CG.
\end{itemize}
Detailed information about the two lattices is shown in table \ref{tab:info}.

	\begin{table*}[th]
		\setstretch{1.5}
		\centering
		\caption{Dirac operators used in the measurement and evolution phases, as well as other information of the two lattices used in this work. HISQ means Highly Improved Staggered Quark. $a^{-1}$ is the inverse lattice spacing of the lattices. d.o.f. refers to the number of real degrees of freedom of the linear system for solving MDWF Dirac equation, and $\kappa$ refers to the condition number of the MDWF Dirac operator with light input quark mass $m_l$.}
        \label{tab:info}
\begin{tabular}{c|c|c|c|c|c|c|c}
\hline
\hline
label & 4D volume & Dirac operator: measurement & Dirac operator: evolution & MDWF d.o.f. & $\kappa$ & $m_l$ & $a^{-1}$ [GeV] \\
\hline
\texttt{RBC96} & $96^3 \times 192$  & MDWF $L_s=12$ & 2+1 flavor MDWF $L_s=12$ & $4.9\times10^{10}$ & $2.7 \times 10^7$ & $0.00054$ & $2.8$ \\
\hline
\texttt{CAL64} & $64^3 \times 96$   & MDWF $L_s=12$ & 2+1+1 flavor HISQ & $7.2\times10^{9}$ & $3.5 \times 10^6$ & $0.00152$ & $2.2$ \\
\hline
\hline
\end{tabular}
	\end{table*}

	\begin{table*}[th]
		\setstretch{1.5}
		\centering
		\caption{MSPCG (with and without reduced-$L_s$ acceleration) is used on Dirac equation ($D^\dagger Dx=y$) linear solves on SUMMIT to a tolerance of $10^{-12}$ on the \texttt{RBC96} lattice. $y$ is a Gaussian random source vector. The times are given in units of seconds. The outer iterations are performed in 16-bit fixed-point format, while the residuals are corrected with reliable updates in double precision. The preconditioning is performed in 16-bit fixed-point format.}
        \label{tab:rbc}
\begin{tabular}{c|c|c|c|c|c|c|c|c}
\hline
\hline
nodes & local volume & solver & inner iteration & $\mu$ & outer iteration & reliable update & time & speedup \\
\hline
\multirow{ 3}{*}{$128$} & \multirow{ 3}{*}{$16\times24\times24\times24$}  & CG    & $-$ & $-$ & $42357$ & $473$ & \colorbox{violet!60}{$783.5$} & $1$ \\
&   & MSPCG & $6$ & $-$ & $14264$ & $166$ & \colorbox{green!60}{$744.5$} & $1.05$ \\
& & MSPCG+reduced-$L_s$ & $8$ & $0.437$ & $14849$ & $173$ & \colorbox{red!60}{$542.2$} & $1.44$\\
\hline
\multirow{ 3}{*}{$256$} & \multirow{ 3}{*}{$16\times24\times12\times24$}  & CG    & $-$ & $-$ & $42342$ & $473$ & \colorbox{violet!60}{$577.6$} & $1$ \\
&   & MSPCG & $6$ & $-$ & $14808$ & $172$ & \colorbox{green!60}{$458.3$} & $1.26$ \\
&   & MSPCG+reduced-$L_s$ & $8$ & $0.437$ & $15693$ & $182$ & \colorbox{red!60}{$381.5$}  & $1.51$ \\
\hline
\hline
\end{tabular}
	\end{table*}

	\begin{table*}[th]
		\setstretch{1.5}
		\centering
		\caption{MSPCG (with and without reduced-$L_s$ acceleration) is used on Dirac equation ($D^\dagger Dx=Dy$)  linear solves on SUMMIT to a tolerance of $10^{-10}$ on the \texttt{CAL64} lattice. $y$ is a random source vector. The times are given in units of seconds. The outer iterations are performed in 16-bit fixed-point format, while the residuals are corrected with reliable updates in double precision. The preconditioning is performed in 16-bit fixed-point format.}
\label{tab:cal}
\begin{tabular}{c|c|c|c|c|c|c|c|c}
\hline
\hline
nodes & local volume & solver & inner iteration & $\mu$ & outer iteration & reliable update & time & speedup \\
\hline
\multirow{ 3}{*}{$16$} & \multirow{ 3}{*}{$32\times32\times16\times16$}  & CG    & $-$ & $-$ & $11383$ & $133$ & \colorbox{violet!60}{$217.1$} & $1$ \\
&   & MSPCG & $6$ & $-$ & $4121$ & $53$ & \colorbox{green!60}{$244.1$} & $0.89$ \\
& & MSPCG+reduced-$L_s$ & $6$ & $0.850$ & $4778$ & $60$ & \colorbox{red!60}{$172.4$} & $1.26$\\
\hline
\multirow{ 3}{*}{$64$} & \multirow{ 3}{*}{$16\times16\times16\times16$}  & CG    & $-$ & $-$ & $11384$ & $133$ & \colorbox{violet!60}{$92.5$} & $1$ \\
&   & MSPCG & $6$ & $-$ & $4464$ & $57$ & \colorbox{green!60}{$90.2$} & $0.98$ \\
&   & MSPCG+reduced-$L_s$ & $6$ & $0.847$ & $5312$ & $66$ & \colorbox{red!60}{$70.6$}  & $1.31$ \\
\hline
\hline
\end{tabular}
	\end{table*}

The results show the reduced-$L_s$ operator, with the trained transform matrix $T$, is able to reduce the outer iteration count as a preconditioner for almost as much as the original operator is able to do. It has little to no effect on the overall convergence of the preconditioned CG.

Note that the transform matrix $T$ is trained with the set of vector
$v_i$ generated from one of the configurations. The trained $T$ is
then used on another configuration: this verifies that the training of
$T$ is independent of which gauge configuration the inversions are
performed on. Once $T$ is trained, it can be used on any other
configurations of the same physical ensemble.

	\section{Conclusion}
	
	We have proven the feasibility of using local preconditioning
        to improve the multi-node scaling of lattice QCD simulations
        with (M\"obius) domain wall fermions. MSPCG is able to reduce
        the (outer) iteration count by a factor of $3$ with as few as
        $6$ inner preconditioning iterations and achieve a $51\%$
        time-to-solution speedup on $256$ SUMMIT nodes utilizing the
        tensor core on the V100 GPUs, together with various other
        optimizations and improvements.  This is the first time the
        tensor cores, which are specifically designed to perform
        matrix-multiply operations, have been used to speed up lattice
        QCD simulations which are traditionally thought of as being
        bandwidth bound (whether memory or network).
	
        Currently the performance of the preconditioning step is
        mainly limited by the GPU memory bandwidth and its cache
        efficiency, these are likely to be improve in the future at a
        faster rate than the the network bandwidth.  Thus, we expect
        MSPCG to bring more significant speedup with this ever growing
        inequality between local and non-local memory bandwidth.  The
        idea of using local preconditioning to improve overall scaling
        can also be applied to lattice QCD simulation with other
        fermion formulations and other scientific computing fields
        where halo communication is the limiting factor for multi-node
        scaling.

        The acceleration that the tensor cores provide, applied
        through MSPCG, demonstrates that new algorithms can become
        feasible that would have otherwise led to a net slowdown.  We
        expect this trend to continue going forward, where increased
        local tensor compute capability at low precision will give rise to
        increased algorithmic innovation to reduce the actual time to
        high-precision science.
	
	\section{Acknowledgments}
	
	The authors wish to thank Duo Guo, Christopher Kelly and Norman Christ for the suggestions and comments, and want to thank Runzhi Wang for suggestion on using the line search method.
	
    The full numerical implementation of algorithm is available in \href{https://github.com/lattice/quda}{\texttt{QUDA}}.
	Numerical experiments that contribute to this work are also done in \texttt{CPS}, \href{https://github.com/paboyle/Grid}{\texttt{Grid}} and \href{https://github.com/waterret/Qlattice}{\texttt{Qlattice}}.

    The \texttt{RBC96} lattice ensembles are generated by RBC/UKQCD collaborations, using the \texttt{CPS} code accelerated by QUDA on the SUMMIT supercomputer at Oak Ridge Leadership Computing Facility at the Oak Ridge National Laboratory, which is supported by the Office of Science of the U.S. Department of Energy.

    R.D.M and J.T (while a student at Columbia University) are supported in part by U.S. Department of Energy contract No. DE-SC0011941. C.J. is supported in part by U.S. Department of Energy contract No. DE-SC0012704.

    The \texttt{CAL64} lattice ensembles are generated by the CalLat collaboration, using the \href{https://github.com/milc-qcd/milc_qcd}{\texttt{MILC}} code accelerated by QUDA on the SUMMIT supercomputer at Oak Ridge Leadership Computing Facility at the Oak Ridge National Laboratory, which is supported by the Office of Science of the U.S. Department of Energy under Contract No. DE-AC05-00OR22725.
	
	\bibliographystyle{ACM-Reference-Format}
	\bibliography{library}
	
\end{document}